%% file: rhorho_preprint.tex
\documentclass[aps,prl,preprint,tightenlines,superscriptaddress,showpacs,byrevtex]{revtex4}

\def\cp{$CP$\/}
\def\mbc{$M^{}_{\rm bc}$}
\def\deltaE{$\Delta E$}

\def\phitwo{$\phi^{}_2$}

\def\meve{~MeV}
\def\gevm{~GeV/$c^2$\/}
\def\gevp{~GeV/$c$\/}
\def\geve{~GeV}

\def\ra{\!\rightarrow\!}
\def\bbar{\overline{B}{}^{\,0}}
\def\dbar{\overline{D}{}^{\,0}}

\def\brhorho{$B^0\ra\rho^+\rho^-$}
\def\brho2pi{$B^0\ra\rho^\pm\pi^\mp\pi^0$}
\def\b4pi{$B^0\ra\pi^+\pi^0\pi^-\pi^0$}

\def\arhorho{${\cal A}$}
\def\srhorho{${\cal S}$}

\def\cone{\cos\theta^{}_1}
\def\ctwo{\cos\theta^{}_2}
\def\conesq{\cos^2\theta^{}_1}
\def\ctwosq{\cos^2\theta^{}_2}

\def\sonesq{\sin^2\theta^{}_1}
\def\stwosq{\sin^2\theta^{}_2}

\usepackage{graphicx} 
\graphicspath{{ps}}

\usepackage{epsf}     
\usepackage{dcolumn}  

\begin{document}

\vspace*{-3\baselineskip}
\resizebox{!}{3cm}{\includegraphics{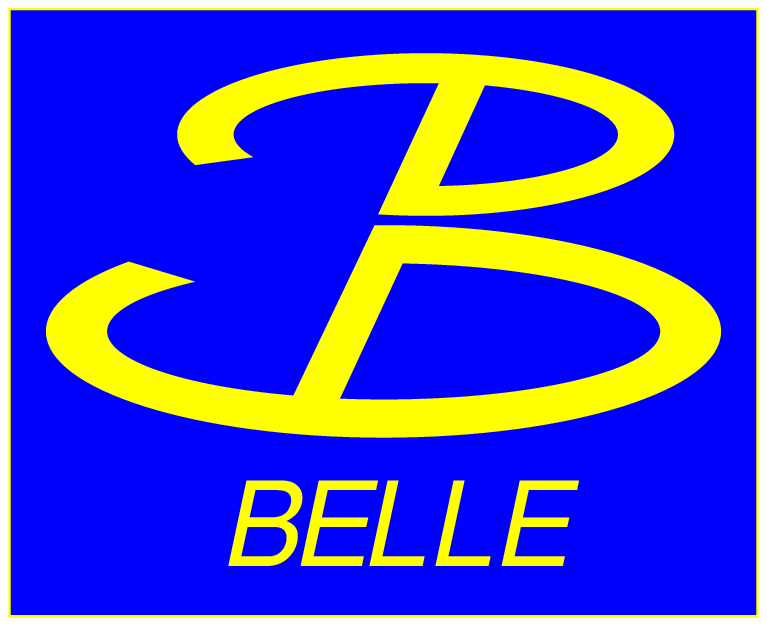}}

\preprint{\vbox{ \hbox{   }
                 \hbox{BELLE-CONF-0545}
                 \hbox{UCHEP-05-02}
}}

\title{ \quad\\[0.5cm]  
{\boldmath Measurement of the branching fraction, polarization, and 
\cp\ asymmetry in \brhorho\ decays} }

\input{author-conf2005.tex}

\noaffiliation

\begin{abstract}
We have measured the branching fraction, longitudinal
polarization fraction $f^{}_L$, and the \cp\ asymmetry 
coefficients ${\cal A}$ and ${\cal S}$ in \brhorho\ decays.
These results are obtained from a $253\,{\rm fb}^{-1}$ data sample 
containing 275 million $B\bar{B}$ pairs 
collected by the Belle detector running at the 
KEKB $e^+ e^-$ collider. We obtain 
$B=\left[\,24.4\,\pm 2.2\,({\rm stat})\,^{+3.8}_{-4.1}\,({\rm syst})\,\right]
\times 10^{-6}$,
$f^{}_L =
0.951\,^{+0.033}_{-0.039}\,({\rm stat})\,^{+0.029}_{-0.031}\,({\rm syst})$,
${\cal A} = 0.00\,\pm 0.30\,({\rm stat})\,^{+0.10}_{-0.09}\,({\rm syst})$, and
${\cal S} = 0.09\,\pm 0.42\,({\rm stat})\,\pm 0.08\,({\rm syst})$.
These values are used to determine the CKM phase angle~\phitwo\ 
via an isospin analysis; the central value and $1\sigma$ error 
are $(87\,\pm 17)^\circ$, and 
$59^\circ < \phi^{}_2 < 115^\circ$ at 90\% CL.
\end{abstract}


\maketitle

\tighten

{\renewcommand{\thefootnote}{\fnsymbol{footnote}}}
\setcounter{footnote}{0}

The decay \brhorho\ has received much attention in the
literature because it allows one to determine the CKM 
phase angle \phitwo~\cite{alpha} with relatively little 
theoretical uncertainty. This is because the penguin 
contribution to \brhorho\ is small, as indicated by the 
small branching fraction for $B^0\ra\rho^0\rho^0$~\cite{babar_rho0rho0}. 
The angle \phitwo\ is determined by measuring the 
decay time ($\Delta t$) distribution of $B^0$ and 
$\bbar$ decays; the difference between the distributions 
is fit to the function ${\cal A}\cos(\Delta m\,\Delta t)\,+\, 
{\cal S}\sin(\Delta m\,\Delta t)$, where
$\Delta m$ is the mass difference 
between the two $B^0$-$\bbar$ mass eigenstates, and the
\cp\ asymmetry coefficients \arhorho\ and \srhorho\ depend 
on~\phitwo. To determine \phitwo\ requires knowledge of the 
polarization of the $\rho$ mesons, as different polarization
amplitudes can give different values of  ${\cal A}$ and ${\cal S}$ 
for the same~\phitwo. For a negligible penguin contribution, 
the sign of ${\cal S}$ for the \cp-odd 
transversity amplitude $A^{}_\perp$ will be opposite to 
that for the \cp-even amplitudes $A^{}_0$ and $A^{}_{\parallel}$.
In this paper we present a measurement of the branching
fraction, longitudinal polarization fraction ($f^{}_L$),
and coefficients \arhorho\ and \srhorho\ for \brhorho\ decays.

The data sample consists of 253~fb$^{-1}$ recorded by the 
Belle experiment running at the KEKB energy-asymmetric $e^+e^-$ 
collider~\cite{kekb}. The Belle detector~\cite{belle_detector}
consists of a silicon vertex detector (SVD), 
a 50-layer central drift chamber (CDC), an array of aerogel 
threshold Cherenkov counters (ACC), time-of-flight 
scintillation counters (TOF), and an electromagnetic 
calorimeter comprised of CsI(Tl) crystals. These detectors 
are located within a superconducting solenoid coil 
providing a 1.5~T magnetic field. An iron 
flux-return located outside the coil is 
instrumented to identify muons and detect 
$K^0_L$ mesons. 
Two inner detector configurations were used: a 2.0 cm beampipe
and a 3-layer silicon vertex detector were used for the first sample
of 152 million $B\overline{B}$ pairs, and a 1.5 cm beampipe, a 4-layer
silicon detector, and a small-cell inner drift chamber were used 
for the remaining 123 million $B\overline{B}$ pairs~\cite{Ushiroda}.  

Candidate \brhorho, $\rho^\pm\ra\pi^\pm\pi^0$ decays 
are selected by requiring two oppositely-charged tracks  
satisfying $p^{}_T>0.10$\gevp, $dr<2.0$~mm, and $|dz|<4.0$~cm, 
where $p^{}_T$ is the momentum transverse to the beam axis, 
and $dr$ and $dz$ are the radial and longitudinal distances, 
respectively, between the track and the beam 
crossing point. The tracks are fit to a common 
vertex, and the resulting $\chi^2$ is required 
to be satisfactory.
We require that the tracks be identified 
as pions based on information from the TOF and ACC systems, 
and from $dE/dx$ measurement in the CDC. This information 
is combined into a likelihood for a track to be 
a pion ($L^{}_\pi$) or a kaon ($L^{}_K$). To identify pions, 
we require that the ratio $L^{}_K/(L^{}_K+L^{}_\pi)$
be less than~0.40. The efficiency of this requirement 
is approximately 89\%, and the kaon misidentification 
rate is about 10\%. Tracks are rejected if they satisfy 
an electron identification criterion based on information 
from the ECL system.

The $\pi^\pm$ candidates are subsequently combined with 
$\pi^0$ candidates. The latter are reconstructed from 
pairs of photons having an invariant mass in the range
0.1178--0.1502\geve, which corresponds to $\pm 3\sigma$ 
in $\pi^0$ mass resolution. The energy of each photon in the 
laboratory frame must be $>50~(90)$\meve\ in the ECL barrel 
(endcap) region, which subtends the angular range with 
respect to the beamline of $32^\circ$--$129^\circ$ 
($17^\circ$--$32^\circ$ and $129^\circ$--$150^\circ$).
In order to reduce combinatorial background, $\pi^0$
candidates are required to have $p>0.35$\gevp\ in the
$e^+e^-$ center-of-mass (CM) frame. 
To identify $\rho^\pm\ra\pi^\pm\pi^0$ decays, we require 
that $m^{}_{\pi^\pm\pi^0}$ be in the range 0.62--0.92\gevm;
this window corresponds to $\pm 2\sigma$ in the $\pi^\pm\pi^0$
mass distribution. To further reduce combinatorial background, 
we require that each $\rho$ candidate satisfy $-0.80<\cos\theta<0.98$, 
where $\theta$ is the angle between the direction of the $\pi^0$ 
and the negative of the $B^0$ momentum in the $\rho^\pm$ 
rest frame.

To select \brhorho\ decays, we calculate the quantities
$M^{}_{\rm bc}\!\equiv\!\sqrt{E^2_{\rm beam}-p^2_B}$ and
$\Delta E\!\equiv\!E^{}_B-E^{}_{\rm beam}$, where $E^{}_B$ 
and $p^{}_B$ are the reconstructed energy and momentum 
of the $B$ candidate, and $E^{}_{\rm beam}$ is the beam 
energy, all evaluated in the CM frame.
The $\Delta E$ distribution has a long tail
on the lower side due to incomplete containment of the 
electromagnetic shower in the ECL. We define a signal region
$5.27~{\rm GeV}/c^2\!<\!M^{}_{\rm bc}\!<\!5.29$\gevm\ and 
$-0.12~{\rm GeV}\!<\!\Delta E\!<\!0.08$\geve. The fraction of events 
having multiple candidates in this region is 9.5\%. Many of
these arise from fake $\pi^0$'s combining with good tracks,
and thus we choose the best candidate 
based on the mass difference $|m^{}_{\gamma\gamma}-m^{}_{\pi^0}|$. 
From MC simulation we find that this criterion correctly identifies 
the \brhorho\ decay 87\% (95\%) of the time for longitudinal (transverse)
polarization. Signal decays that have one or more daughters 
incorrectly identified but nonetheless pass all selection 
requirements are referred to as ``self-cross-feed'' (SCF) 
background. Most such decays have the $\pi^0$ daughter swapped 
with a $\pi^0$ originating from the rest of the event; this 
preserves the vertex position and thus the $\Delta t$ value.

We identify whether a $B^0$ or $\bbar$ evolved and 
decayed to $\rho^+\rho^-$ by tagging the $b$ flavor of 
the non-signal (opposite-side) $B$ decay in the event. 
This is done using a tagging algorithm~\cite{tagging} 
that categorizes charged leptons, kaons, and $\Lambda$'s 
found in the event. The algorithm returns two parameters: 
$q$, which equals $+1\,(-1)$ when the opposite-side $B$
is likely to be a $B^0\,(\bbar)$; and $r$, which 
represents the quality of the tag as determined
from MC simulation and varies from $r\!=\!0$ for 
no flavor discrimination to $r\!=\!1$ for unambiguous 
flavor assignment. 

The dominant background is $e^+e^-\ra q\bar{q}\ (q=u,d,s,c)$
continuum production. We discriminate against this using 
event topology: $e^+e^-\ra q\bar{q}$ processes in the CM 
frame tend to be jet-like, while $e^+e^-\ra B\overline{B}$ 
events tend to be 
spherical. To quantify this sphericity, we calculate a set 
of 16 modified Fox-Wolfram moments and combine them into a Fisher 
discriminant~\cite{KSFW}. We calculate a probability density 
function (PDF) based on this discriminant and multiply it 
by a PDF for $\cos\theta^{}_B$, where $\theta^{}_B$ is the 
polar angle in the CM frame between the $B$ direction and 
the positron beam direction. $B\overline{B}$ events are produced 
with a $1-\cos^2\theta^{}_B$ distribution while continuum events 
are produced uniformly in $\cos\theta^{}_B$. The PDFs for signal 
and continuum background are obtained using MC simulation and the 
data sideband $5.21~{\rm GeV}/c^2\!<\!M^{}_{\rm bc}\!<\!5.26$\gevm, 
respectively. We use the products of the PDFs to calculate a 
signal likelihood ${\cal L}^{}_s$ and continuum likelihood 
${\cal L}^{}_{q\bar{q}}$ and require that 
${\cal R} ={\cal L}^{}_s/({\cal L}^{}_s + {\cal L}^{}_{q\bar{q}})$
be above a threshold. This threshold is determined by optimizing a 
figure-of-merit $N^{}_S/\sqrt{N^{}_S+N^{}_B}$, where 
$N^{}_S$ ($N^{}_B$) is the number of signal (background) 
events estimated to be in the signal region from MC simulation 
(extrapolating from an \mbc\ sideband). For this calculation we 
assume a \brhorho\ branching fraction of $25\times 10^{-6}$.
As the tagging parameter $r$ also discriminates against 
$q\bar{q}$ background, we divide the data into six $r$ 
intervals (denoted by $\ell\!=\!1\!-\!6$) and determine the 
${\cal R}$ threshold separately for each. 
The ${\cal R}$ requirement removes about 97\% of continuum 
events while retaining about 62\% of \brhorho\ signal. The efficiency 
of all selection criteria as determined from MC simulation is 
$(2.19\pm 0.02)$\%. This value corresponds to $f^{}_L\!=\!1$;
the efficiency for our measured value of $f^{}_L$ (which is 
slightly less than one --\,see below) is about 4\% higher, and 
we include this difference as an additional systematic error in
the branching fraction.

We determine the signal yield in two steps: we first fit 
the \mbc-\deltaE\ distribution to obtain the fraction of 
\brhorho\,+\,$B^0\ra\rho^\pm\pi^\mp\pi^0$ non-resonant decays 
(as this fit cannot distinguish between the two modes); we 
then fit the $m^{}_{\pi^\pm\pi^0}$ distribution to determine 
the non-resonant fraction and hence the $\rho^+\rho^-$ signal
yield.

The first fit is an unbinned maximum likelihood (ML) fit 
to the two-dimensional \mbc-\deltaE\ distribution in the 
wide range $5.21~{\rm GeV}/c^2\!<\!M^{}_{\rm bc}\!<\!5.29$\gevm\ 
and $-0.20~{\rm GeV}\!<\!\Delta E\!<\!0.30$\geve. The fit includes 
components for the signal and several backgrounds:
continuum events, $b\ra c$ decays, and $b\ra u$
decays. The PDFs for signal and $b\ra u$ background
are modeled by smoothed two-dimensional histograms 
obtained from large MC samples. The PDF for
$b\ra c$ background is the product of 
a threshold (``ARGUS''~\cite{argus}) function
for \mbc\ and a quadratic polynomial for \deltaE,
also obtained from MC simulation.
The PDF for continuum background is taken to be
an ARGUS function for \mbc\ and a linear function 
for \deltaE; the slope of the linear function 
depends on the tag quality ($r$) bin~$\ell$, and all seven 
shape parameters are floated in the fit. The signal PDF 
is adjusted to account for small differences 
observed between data and MC for a high-statistics
sample of $B^+\ra\dbar\rho^+,\,\dbar\ra K^+\pi^-\pi^0$ decays,
which, like the signal mode, contain two neutral pions.

\begin{figure}[t]
\centerline{\epsfxsize=6.0in \epsfbox{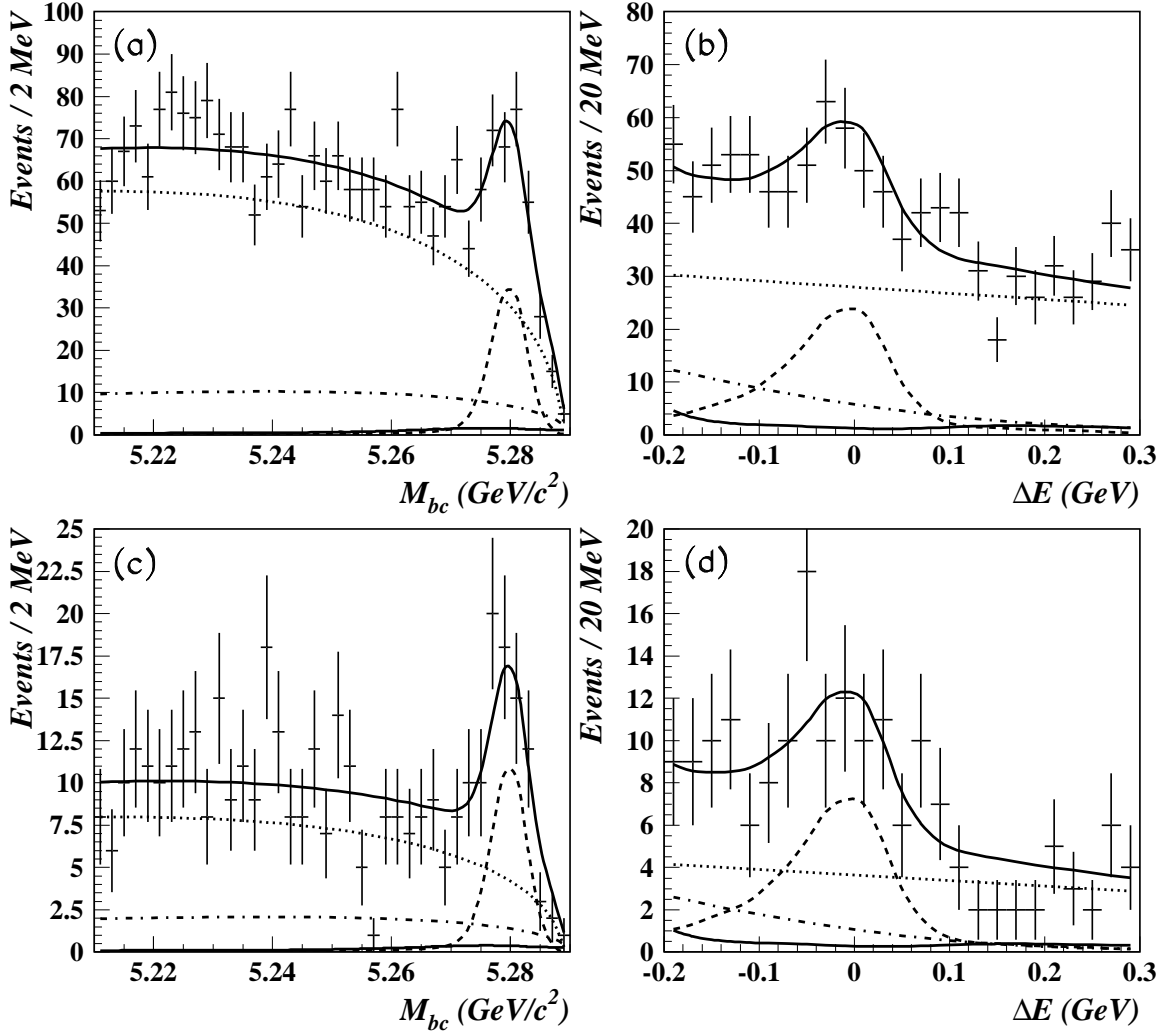}}
\caption{Left: projection in \mbc\ for events satisfying
$-0.10~{\rm GeV}\!<\!\Delta E\!<\!0.06$\geve. Right: projection 
in \deltaE\ for events satisfying 
$5.273~{\rm GeV}/c^2\!<\!M^{}_{\rm bc}\!<\!5.290$\gevm. 
The top plots correspond to all $r$ values, and the bottom 
plots correspond to $0.75\!<\!r\!<\!1.0$.
The curves show projections of the ML fit result: the 
dashed curve represents $\rho^+\rho^- +\rho\pi\pi$ events,
the dotted curve represents $q\bar{q}$ background, 
the dot-dashed curve represents $b\ra c$ background, 
the small solid curve represents $b\ra u$ background, 
and the large solid curve shows the overall result. }
\label{fig:one}
\end{figure}

The $b\ra u$ background is dominated by $B^{(0,\,+)}\ra \rho\,\pi$,
$B\ra a^{}_1\pi$, and $B\ra a^{}_1\,\rho$ decays. As their
contributions are small, their normalization is fixed to 
the yield obtained from MC simulation. For
$B^+\ra a^+_1\pi^0$ and $B\ra a^{}_1\,\rho$, the
branching fractions are unknown; we therefore estimate 
them to be $3\times 10^{-6}$ and $2\times 10^{-6}$, respectively,
and vary these values by 50\% and 100\%, respectively, to 
obtain the systematic error due to these estimates. There 
are thus nine free parameters in the fit, and the resulting 
fraction of \brhorho\,+\,non-resonant decays in the signal 
region is~$(3.0\,\pm 0.4)\%$.
Figure~\ref{fig:one} shows the final event sample and projections 
of the fit result. The $\chi^2$ of the projections divided by the
number of bins is 0.97 for \mbc\ and 1.02 for \deltaE. 

\begin{figure}[t]
\begin{center}
\centerline{\epsfxsize=4.75in \epsfbox{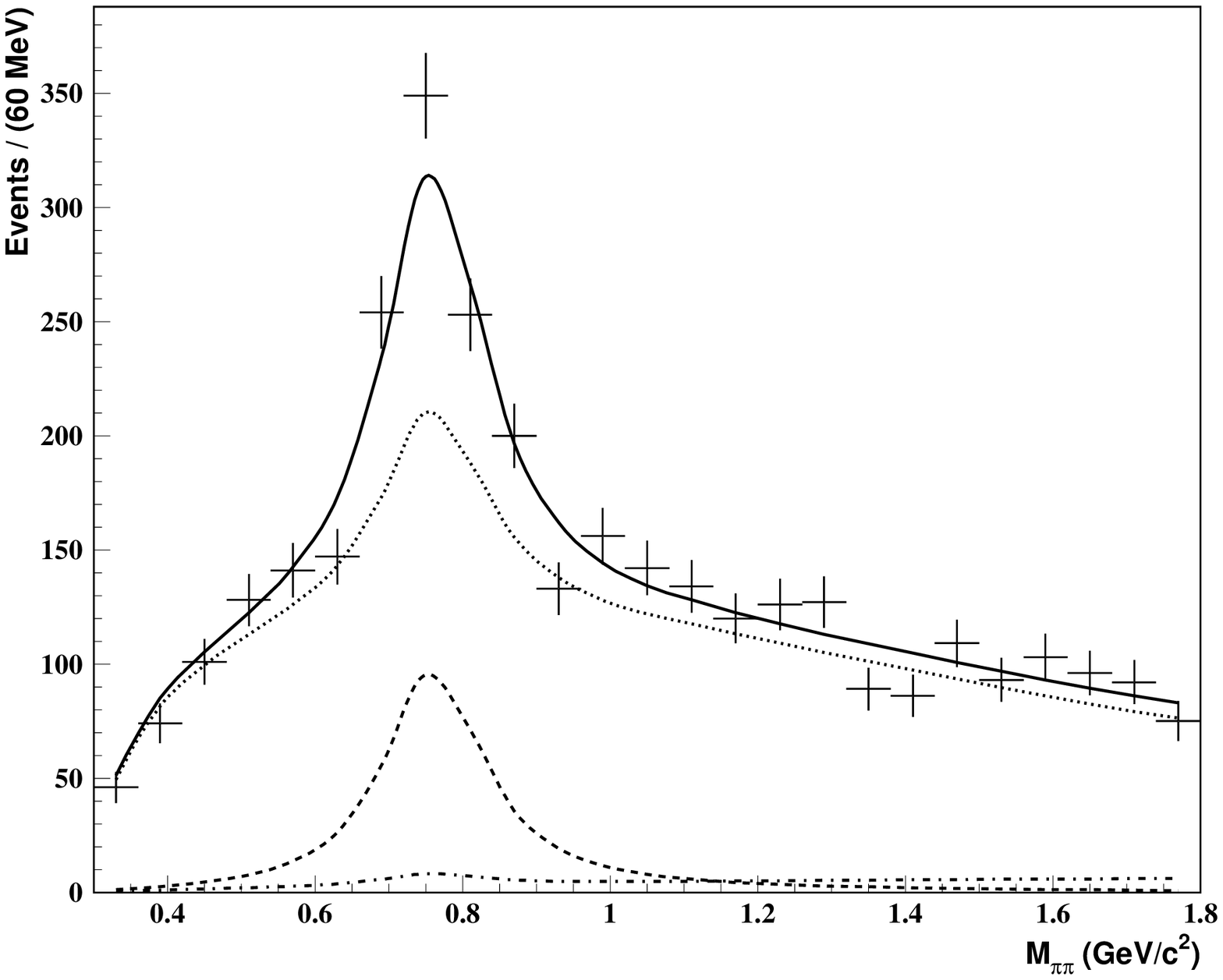}}
\end{center}
\vskip-0.25in
\caption{The $m^{}_{\pi^+\pi^0}$ plus $m^{}_{\pi^-\pi^0}$
distribution (two entries per event) for events satisfying
$5.27~{\rm GeV}/c^2\!<\!M^{}_{\rm bc}\!<\!5.29$\gevm\ and 
$-0.12~{\rm GeV}\!<\!\Delta E\!<\!0.08$\geve. Superimposed
are projections of the ML fit result. The dashed curve
represents $\rho^+\rho^-$ signal; 
the dot-dashed curve represents non-resonant $\rho\pi\pi$ decays; 
the dotted curve represents $q\bar{q}\,+(b\ra c)+(b\ra u)$ backgrounds; 
and the solid curve shows the overall result.}
\label{fig:two}
\end{figure}

To distinguish \brhorho\ decays from non-resonant 
\brho2pi\ and \b4pi\ decays, we do an unbinned ML fit 
to the $\pi^\pm\pi^0$ mass distribution. For this fit we require 
$5.27~{\rm GeV}/c^2\!<\!M^{}_{\rm bc}\!<\!5.29$\gevm\ and  
$-0.12~{\rm GeV}\!<\!\Delta E\!<\!0.08$\geve\ and fit the 
$m^{}_{\pi^\pm\pi^0}$ distribution in the range 0.30--1.80\gevm. 
Only one $\rho$ candidate is required to satisfy 
$0.62~{\rm GeV}/c^2\!<\!m^{}_{\pi^\pm\pi^0}\!<\!0.92$\gevm; 
the mass of the other $\rho$ candidate is then fit.
We include additional PDFs for 
non-resonant $B\ra\rho\pi\pi$ and $B\ra\pi\pi\pi\pi$
components; however, the fit result for the latter
is negligibly small ($\ll\!1\%$) and consistent 
with zero, and thus we set this fraction to zero. 
The PDFs for continuum and $b\ra c$ backgrounds 
are grouped together and taken from the data sideband 
$5.22~{\rm GeV}/c^2\!<\!M^{}_{\rm bc}\!<\!5.26$\gevm;
we use MC simulation to check that the shapes of these 
backgrounds and their ratio in the sideband region are 
very close to those in the signal region. We impose the 
constraint that the fraction of signal\,+\,non-resonant 
events in the $m^{}_{\pi^\pm\pi^0}$ range 0.62--0.92\gevm\ 
equals that which we obtained from the \mbc-\deltaE\ fit;
there is then only one free parameter in the fit ($f^{}_{\rho\pi\pi}$). 
The result of the fit is $f^{}_{\rho\pi\pi}=(8.9\pm 6.2)\%$, 
and thus $N^{}_{\rho\rho}=142\pm 13$ events. Figure~\ref{fig:two} 
shows the data (two entries per event) and the projection 
of the fit result. The $\chi^2$ of the projection divided
by the number of bins is~1.35.

The branching fraction is evaluated as
\begin{eqnarray}
B(B^0\ra\rho^+\rho^-) & = & \frac{N^{}_{\rho\rho}}
{\varepsilon\cdot\varepsilon^{}_{\rm PID}\cdot N^{}_{B\overline{B}}}\,,
\label{eqn:br}
\end{eqnarray}
where $N^{}_{\rho\rho}$ is the number of \brhorho\ candidates
($142\pm 13$),
$N^{}_{B\overline{B}}$ is the number of $B\overline{B}$ pairs
produced [$(274.8\,\pm2.3)\times 10^6$], 
$\varepsilon$ is the geometric acceptance and efficiency
of selection criteria obtained from MC simulation
[$(2.19\pm0.02)\%$], and 
$\varepsilon^{}_{\rm PID}$ is a correction factor for the
pion identification requirement to account for small differences
between data and MC ($0.969\pm 0.012$). 
Inserting these values into Eq.~(\ref{eqn:br}) gives a branching 
fraction of $(24.4\pm 2.2)\times 10^{-6}$, where the error given
is statistical.

There are several sources of systematic uncertainty; their effect 
is evaluated by varying the relevant parameter(s) by $\pm 1\sigma$
and noting the resulting change in the branching fraction.
The different sources and the corresponding changes are:
track reconstruction efficiency (1.2\% per track);
$\pi^0$ reconstruction efficiency (4\% per $\pi^0$);
the $\pi^\pm$ identification efficiency 
$\varepsilon^{}_{\rm PID}$ in Eq.~(\ref{eqn:br}) (1.4\%);
MC statistics used to calculate the acceptance 
$\varepsilon$ in Eq.~(\ref{eqn:br}) (1.0\%); 
the dependence of the acceptance upon $f^{}_L$ (+0.0\%, $-4.1$\%); 
the continuum suppression requirement (13\%); 
and the fraction of signal\,+\,non-resonant decays used 
as a constraint in the $m^{}_{\pi^\pm\pi^0}$ fit.
This last error itself has several components:
the calibration factors used to correct the
\mbc-\deltaE\ distribution of MC \brhorho\ 
events (to better match the data); the \mbc-\deltaE\ 
shapes used for $b\ra c$ background; the 
fraction of $b\ra u$ background, which is obtained 
from MC simulation and includes several estimated 
branching fractions; and the \deltaE\ range used 
in the fit. These components give an overall 
variation in $f^{}_{\rho\rho}\!+f^{}_{\rho\pi\pi}$ 
of $(+2.9\%,-8.6\%)$. We subsequently redo the 
$m^{}_{\pi^\pm\pi^0}$ fit, varying $f^{}_{\rho\rho}\!+f^{}_{\rho\pi\pi}$ 
by this amount; the resulting change in $N^{}_{\rho\rho}$ 
is $(+2.0\%, -6.4\%)$. Combining this in quadrature with 
the other systematic errors gives an overall 
systematic error of $(+16\%,-17\%)$. Thus the 
final result for the branching fraction is
\begin{eqnarray}
B(B^0\ra\rho^+\rho^-) & = &
\left[\,24.4\,\pm 2.2\,({\rm stat})\,^{+3.8}_{-4.1}\,({\rm syst})\,\right]
\times 10^{-6}\,.
\label{eqn:result}
\end{eqnarray}

To determine the polarization of \brhorho\ decays, we do an unbinned
ML fit to the helicity angle distribution $F(\cone,\ctwo)$, where
$\theta^{}_1$ ($\theta^{}_2$) is the angle between the direction 
of the $\pi^0$ from the $\rho^+$ ($\rho^-$) and the negative of
the $B^0$ momentum in the $\rho^+$ ($\rho^-$) rest frame.
For a longitudinal polarization fraction $f^{}_L$, this
distribution is
\begin{eqnarray}
\frac{9}{4}\left[\,f^{}_L\conesq\ctwosq\,+\,\frac{1}{4}
  (1-f^{}_L)\sonesq\stwosq\,\right] .
\label{eqn:angular}
\end{eqnarray}
In the fit, this PDF is multiplied by a two-dimensional 
acceptance function for $(\cone, \ctwo)$
determined from MC simulation. The acceptance is modeled
as the product $A(\cone)\!\cdot\!A(\ctwo)$, where $A$ is 
a polynomial function.

\begin{figure}[t]
\begin{center}
\centerline{\epsfxsize=4.75in \epsfbox{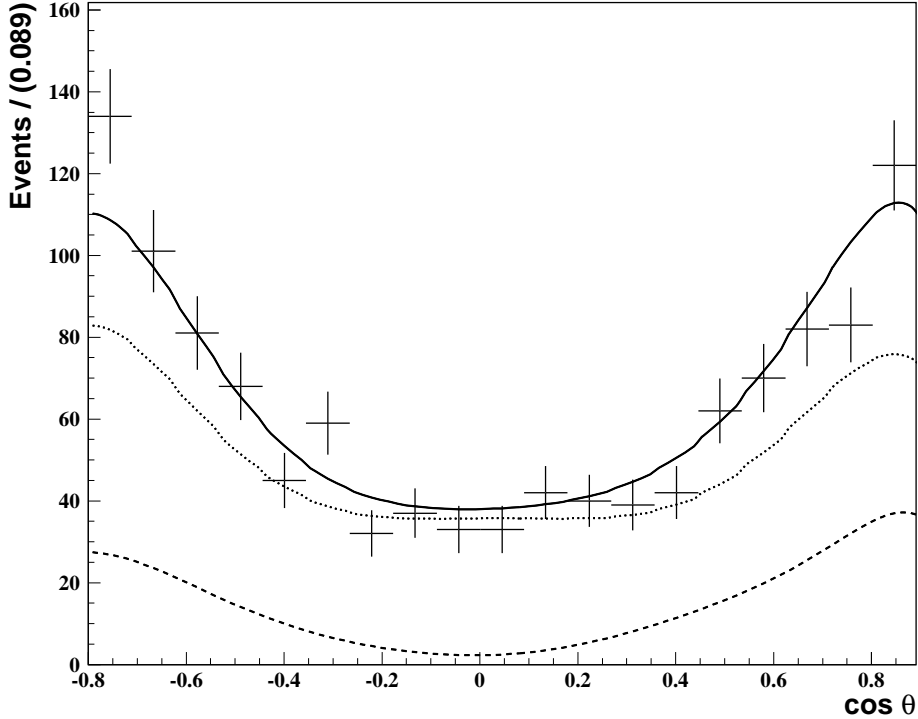}}
\end{center}
\vskip-0.30in
\caption{The $\cos\theta^{}_1$ plus $\cos\theta^{}_2$
distribution (two entries per event) for events satisfying
$5.27~{\rm GeV}/c^2\!<\!M^{}_{\rm bc}\!<\!5.29$\gevm,
$-0.12~{\rm GeV}\!<\!\Delta E\!<\!0.08$\geve, and
$0.62~{\rm GeV}/c^2\!<\!m^{}_{\pi^\pm\pi^0}\!<\!0.92$\gevm. 
Superimposed are projections of the ML fit result. The dashed 
curve represents $\rho^+\rho^-$ signal; the dotted curve
represents non-resonant $\rho\pi\pi$ decays and
$q\bar{q}\,+(b\ra c)+(b\ra u)$ backgrounds, and 
the solid curve shows the overall result. }
\label{fig:three}
\end{figure}

We fit events satisfying 
$5.27~{\rm GeV}/c^2\!<\!M^{}_{\rm bc}\!<\!5.29$\gevm,
$-0.12~{\rm GeV}\!<\!\Delta E\!<\!0.08$\geve, and
$0.62~{\rm GeV}/c^2\!<\!m^{}_{\pi^\pm\pi^0}\!<\!0.92$\gevm.
The likelihood function includes PDFs for signal,
$\rho\pi\pi$ non-resonant decays, and continuum,
$b\ra c$, and $b\ra u$ backgrounds.
The PDF for non-resonant decays is taken to
be constant, and the PDF for $b\ra u$ background 
is taken from MC simulation. The PDFs for continuum 
and $b\ra c$ backgrounds are combined and determined 
from the data sideband 
$5.21~{\rm GeV}/c^2\!<\!M^{}_{\rm bc}\!<\!5.26$\gevm,
$-0.12~{\rm GeV}\!<\!\Delta E\!<\!0.12$\geve;
we use MC simulation to check that the shapes 
of these backgrounds and their ratio in the sideband 
region are very close those in the signal region.
The fraction of signal\,+\,non-resonant decays 
is taken from the previous \mbc-\deltaE\ fit;
the component $f^{}_{\rho\pi\pi}$ is taken from the 
previous $m^{}_{\pi^\pm\pi^0}$ fit.
The fraction of $b\ra u$ background is small and is taken
from MC simulation. Since $f^{}_{(q\bar{q}\,+\,b\rightarrow c)}=
1-f^{}_{\rho\rho}-f^{}_{\rho\pi\pi}-f^{}_{b\rightarrow u}$,
there is only one free parameter in the fit~($f^{}_L$).
The result of the fit is $f^{}_L =0.951\,^{+0.033}_{-0.039}$,
where the error given is statistical. Figure~\ref{fig:three} 
shows the data and projections of the fit result. The $\chi^2$ 
of the fit projections divided by the number of bins is
0.83 for $\cone$ and 1.05 for $\ctwo$.

For this measurement there are eight sources of 
systematic error; their effect is evaluated by varying 
the relevant parameter(s) by $\pm 1\sigma$ and noting the 
resulting change in $f^{}_L$. The different sources and 
the corresponding changes are: 
the signal\,+\,non-resonant fraction $(+0.013, -0.012)$;
the non-resonant fraction $(+0.021, -0.020)$;
the pion identification efficiency, which has a small
effect upon the acceptance $(+0.000, -0.004)$;
misreconstructed \brhorho\ decays $(+0.005, -0.000)$;
the continuum suppression requirement $(\pm 0.013)$; 
possible interference with an $L\!=\!0$ $\pi^\pm\pi^0$
system produced in non-resonant $\rho^\pm\pi^\mp\pi^0$ decays $(\pm 0.005)$;  
a very small bias in the fitting procedure measured from
a large toy MC sample $(+0.000, -0.005)$; and uncertainty 
in the continuum$\,+(b\ra c)$ background shape $(+0.004, -0.014)$. 
This last uncertainty is evaluated by taking the background
shape from alternative \mbc\ and \deltaE\ sidebands.
Combining all systematic errors in quadrature gives 
an overall systematic error of $(+0.029, -0.031)$. 
Thus the final result for the fraction of longitudinally
polarized decays is
\begin{eqnarray}
f^{}_L & = &
0.951\,^{+0.033}_{-0.039}\,({\rm stat})\,^{+0.029}_{-0.031}\,({\rm syst})\,.
\label{eqn:pol_result}
\end{eqnarray}

To fit for the \cp\ asymmetry coefficients \arhorho\ and \srhorho, 
we divide the data into $q\!=\!+1$ and $q\!=\!-1$ tagged decays 
and do an unbinned ML fit to the respective $\Delta t$ distributions. 
At KEKB, the $\Upsilon(4S)$ is produced with a Lorentz boost of 
$\beta\gamma=0.425$ nearly along the electron beamline~($z$);
since the $B^0$ and $\bbar$ mesons are approximately at 
rest in the $e^+e^-$ CM frame, $\Delta t$ is determined 
from the displacement in $z$ between the $f_{CP}$ and 
$f_{\rm tag}$ decay vertices:
$\Delta t \approx (z_{CP} - z_{\rm tag})/\beta\gamma c$.

The likelihood function for event $i$ is
\begin{eqnarray}
{\cal L}^{}_i & = & 
f^{(i)}_{\rho\rho}\,{\cal P}(\Delta t)^{}_{\rho\rho}\ +\  
f^{(i)}_{\rm SCF}\,{\cal P}(\Delta t)^{}_{\rm SCF}\ +\
f^{(i)}_{\rho\pi\pi}\,{\cal P}(\Delta t)^{}_{\rho\pi\pi}\ +\  \\
 & & \hskip0.20in
f^{(i)}_{b\rightarrow c}\,{\cal P}(\Delta t)^{}_{b\rightarrow c}\ +\  
f^{(i)}_{b\rightarrow u}\,{\cal P}(\Delta t)^{}_{b\rightarrow u}\ +\  
f^{(i)}_{q\bar{q}}\,{\cal P}(\Delta t)^{}_{q\bar{q}}\,,  \nonumber
\end{eqnarray}
where 
the event weights $f^{(i)}$ are functions of \mbc\ and \deltaE\
and are normalized to the fractions of events obtained from the 
previous \mbc-\deltaE\ and $m^{}_{\pi^\pm\pi^0}$ fits. The PDFs
${\cal P}(\Delta t)$ for $b\ra c$ and $b\ra u$ backgrounds 
are determined from MC simulation, and the PDF for continuum
$q\bar{q}$ background is determined from an \mbc\ sideband. 
We include an additional PDF for SCF background in which a 
$\pi^\pm$ daughter is swapped with a track from the rest of the
event; this function is determined from MC simulation and is found
to be exponential with an effective lifetime of about~1~ps.
The weighting function $f^{}_{\rm SCF}$ is also determined 
from MC simulation; its normalization is~5.7\% of all 
$\rho^+\rho^-$ candidates.

\begin{figure}[t]
\centerline{\epsfxsize=6.0in \epsfbox{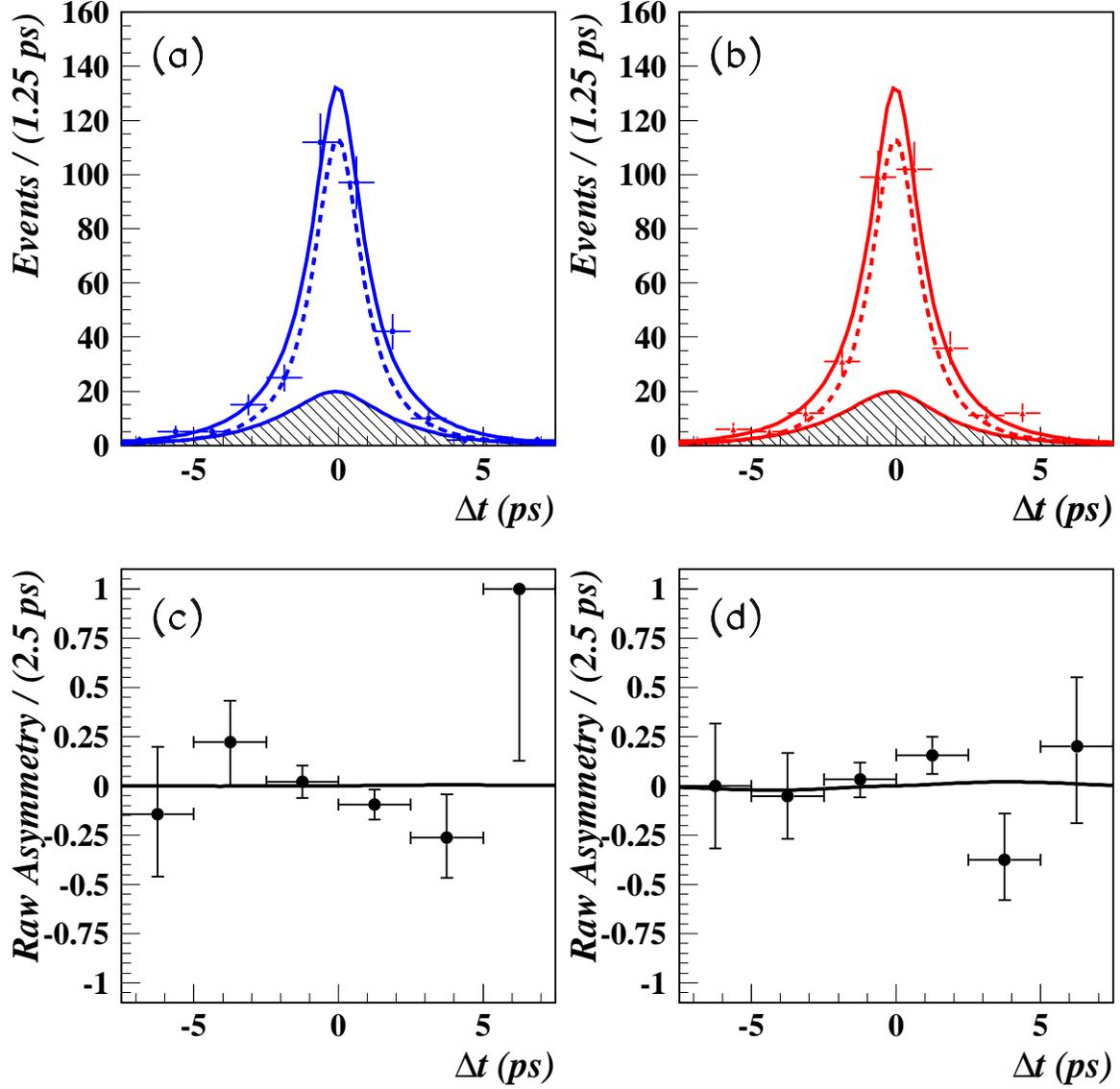}}
\caption{The $\Delta t$ distribution for events satisfying 
$5.27~{\rm GeV}/c^2\!<\!M^{}_{\rm bc}\!<\!5.29$\gevm,
$-0.12~{\rm GeV}\!<\!\Delta E\!<\!0.08$\geve, and
$0.62~{\rm GeV}/c^2\!<\!m^{}_{\pi^\pm\pi^0}\!<\!0.92$\gevm.
Superimposed are projections of the ML fit result.
{\bf (a)}\ $q\!=\!+1$ tags;
{\bf (b)}\ $q\!=\!-1$ tags;
{\bf (c)}\ raw \cp\ asymmetry for $0.0\!<\!r\!<\!0.5$; 
{\bf (d)}\ raw \cp\ asymmetry for $0.5\!<\!r\!<\!1.0$. 
For {\bf (a)} and {\bf (b)}, the hatched distribution
represents signal events and the dashed curve represents
background.
}
\label{fig:four}
\end{figure}

The PDF ${\cal P}^{}_{\rho\rho}(\Delta t)$ is given by
\begin{eqnarray}
\int_{-\infty}^{+\infty}\frac{e^{-|\Delta t'|/\tau^{}_{B^0}}}{4\tau^{}_{B^0}}
\,\biggl\{1-q\,\Delta\omega^{}_{\ell\,(i)} + q(1-2\omega^{}_{\ell\,(i)}) & 
\times & \\
 & & \hskip-2.2in \left[\,{\cal A}\cos(\Delta m\,\Delta t')\ +\ 
{\cal S}\sin(\Delta m\,\Delta t')\,\right]\biggr\}\,
R(\Delta t^{(i)}\!,\Delta t')\,d\Delta t'\,,
\nonumber
\end{eqnarray}
where $R(\Delta t,\Delta t')$ is a resolution function determined 
from data, $\omega^{}_{\ell}$ is the mistag probability for the 
$\ell$th bin of the tagging parameter~$r$, and $\Delta\omega^{}_\ell$
is a possible difference in $\omega^{}_\ell$ between $q\!=\!+1$ 
and $q\!=\!-1$ tags. The PDF ${\cal P}^{}_{\rho\pi\pi}$ is
taken to be exponential with $\tau=\tau^{}_B$ and is
smeared by the same resolution function $R$.
We determine \arhorho\ and \srhorho\ by maximizing
the log-likelihood $\sum_i \log{\cal L}^{}_i$, where $i$
runs over the 656 candidate events satisfying
$5.27~{\rm GeV}/c^2\!<\!M^{}_{\rm bc}\!<\!5.29$\gevm,
$-0.12~{\rm GeV}\!<\!\Delta E\!<\!0.08$\geve, and
$0.62~{\rm GeV}/c^2\!<\!m^{}_{\pi^\pm\pi^0}\!<\!0.92$\gevm.
The results of the fit are ${\cal A} =0.00\,\pm 0.30$
and ${\cal S} =0.09\,\pm 0.42$, where the errors 
given are statistical. These values are consistent
with the no-\cp-violation case ${\cal A}={\cal S}=0$, and
the errors are consistent with expectations based on MC studies.
Figure~\ref{fig:four} shows the data and projections of the fit 
result for $0.0\!<\!r\!<\!0.5$ and $0.5\!<\!r\!<\!1.0$ subsamples; 
the $\chi^2$ of the projections divided by the number of 
bins is 0.98 and 0.97, respectively, which is satisfactory.

For this measurement there are twelve sources of systematic 
error; their effect is evaluated by varying the relevant 
parameter(s) by $\pm 1\sigma$ and noting the resulting change 
in \arhorho\ and \srhorho. The different sources and the 
corresponding changes are listed in Table~\ref{tab:cpv_systematics}. 
The error due to wrong-tag fractions is evaluated by
varying $\omega^{}_\ell$ and $\Delta\omega^{}_\ell$ values.
The error due to $b\ra c$ and continuum $\Delta t$
distributions is evaluated by varying the $\sim$\,40 
parameters used to describe these distributions.
The error due to the component fractions is evaluated
by varying the fractions obtained from the \mbc-\deltaE\ fit.
The error due to self-cross-feed (SCF) events is evaluated by varying 
both the SCF fraction (by 50\%) and the effective lifetime used for 
SCF events (by $2\sigma$). The effect of a \cp\ asymmetry in 
$b\ra c$ and $q\bar{q}$ backgrounds is evaluated by adding such 
asymmetries to the $b\ra c$ and $q\bar{q}$ $\Delta t$ distributions; 
the sizes of these asymmetries are taken to be those 
obtained from fitting an \mbc\ sideband in the data
(these values are small and consistent with zero). 
The error due to a possible fitting bias is evaluated using 
a full MC simulation, and the error due to tag-side
interference~\cite{Long} is evaluated using a toy MC simulation. 
The error due to the transversity amplitudes $A^{}_\perp$ and 
$A^{}_\parallel$ (which may have different values of \arhorho\ 
and \srhorho) is evaluated in two steps:
first setting $f^{}_L$ equal to its central value and varying
the values of (\arhorho, \srhorho) for $A^{}_\perp$ and $A^{}_\parallel$
over the range $-1$ to $+1$; second, using common values of
(\arhorho, \srhorho) for all transversity amplitudes (except 
for an extra minus sign for $A^{}_\perp$) and varying~$f^{}_L$.
Combining all systematic errors in quadrature gives overall 
systematic errors of $(+0.10, -0.09)$ for ${\cal A}$ and 
$\pm\,0.08$ for ${\cal S}$. Thus the final result is
\begin{eqnarray}
{\cal A} & = &
0.00\,\pm 0.30\,({\rm stat})\,^{+0.10}_{-0.09}\,({\rm syst}) \\
 & & \nonumber \\
{\cal S} & = &
0.09\,\pm 0.42\,({\rm stat})\,\pm 0.08\,({\rm syst})\,.
\label{eqn:cpv_result}
\end{eqnarray}

\begin{table}[htb]
\caption{ Systematic errors for the \cp\ asymmetry
coefficients ${\cal A}$ and ${\cal S}$. }
\label{tab:cpv_systematics}
\vskip0.20in
\begin{tabular}{|l|c|c|c|c|}
\hline
{\bf Type} & \multicolumn{2}{|c|}{\boldmath $\delta {\cal A}$ ($\times 10^{-2}$)}   
     & \multicolumn{2}{|c|}{\boldmath $\delta {\cal S}$ ($\times 10^{-2}$)} \\\cline{2-5}
            &  {\boldmath $+\sigma$} & {\boldmath $-\sigma$} & 
{\boldmath $+\sigma$} & {\boldmath $-\sigma$} \\
\hline

Wrong tag fractions   &  0.5  &  0.6   &   0.8  &  0.8  \\
Parameters $\Delta m,\,\tau^{}_{B^0}$  
                      &  0.1  &  0.1   &   0.9  &  0.9  \\
\hline
Resolution function   &  1.3  &  1.3   &   1.3  &  1.3  \\
Background $\Delta t$ distributions    &   1.6  &  1.5  &  2.3  & 2.5 \\
\hline
Component fractions       &  2.1  &  2.6   &   5.0  &  4.5  \\
$\rho\pi\pi$ non-resonant fraction &  0.0 & 0.0  &  0.6  & 0.6 \\
Self-cross-feed (SCF) fraction     &  0.0 & 0.0  &  0.2  & 0.2 \\
\hline  
Background asymmetry  &  0.0      &  2.0   &   0.0       &  4.3  \\
\hline
Possible fitting bias    &  0.0      &  0.2   &    0.6  &  0.0   \\        
\hline
Vertexing             &  4.1  &  2.8   &    0.4  &  0.4   \\
\hline
Tag-side interference  &  7.2 &  5.5 &   0.3  & 0.0 \\
\hline
$f^{}_\perp,\,f^{}_\parallel$ (transversely- &  &  &  & \\
\ \ \ polarized components)   & 5.2  & 4.5  & 6.0  & 4.7    \\
\hline
{\bf Total}     &  $+10.2$\ \ & $-8.6$\ \  &  $+8.4$\ \ &  $-8.4$\ \  \\  
\hline
\end{tabular}
\end{table}

\begin{figure}[t]
\begin{center}
\centerline{\epsfxsize=12.5cm \epsfbox{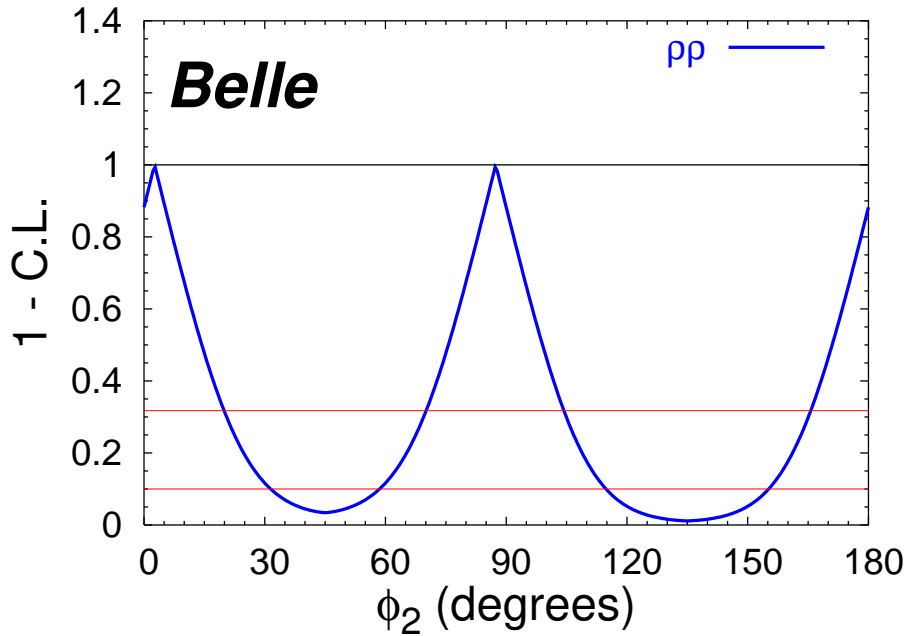}}
\end{center}
\vskip-0.30in
\caption{Results of fitting the branching fractions
$B(B^0\ra\rho^+\rho^-),\,B(B^+\ra\rho^+\rho^0)$,
$B(B^0\ra\rho^0\rho^0)$, and the $\rho^+\rho^-$ 
asymmetry parameters \arhorho\ and \srhorho\ for 
the angle~\phitwo, using the isospin relations
of Ref.~\cite{gronau_london}. The vertical axis
is one minus the confidence level (see text). 
The horizontal line at $1\!-\!{\rm CL}\!=\!0.317$ 
corresponds to a 68.3\% CL interval for \phitwo, 
and the horizontal line at $1\!-\!{\rm CL}\!=\!0.10$ 
corresponds to a 90\% CL interval.}   
\label{fig:phi2constraint}
\end{figure}

We use these values along with the measured branching 
fraction for \brhorho\ and previously-measured branching 
fractions for $B^+\ra\rho^+\rho^0$~\cite{pdg} and 
$B^0\ra\rho^0\rho^0$~\cite{babar_rho0rho0} to constrain
the angle~\phitwo. We use the isospin relations of 
Ref.~\cite{gronau_london} (originally 
applied to the $B\ra\pi\pi$ system), neglecting a possible
$I\!=\!1$ contribution to the \brhorho\ amplitude~\cite{falk}. 
We first fit the measurements to obtain a minimum $\chi^2$; this 
is denoted $\chi^2_{\rm min}$. We then scan values of \phitwo\ from 
0$^\circ$--180$^\circ$ and for each value calculate the difference 
$\Delta\chi^2\equiv\chi^2(\phi^{}_2)-\chi^2_{\rm min}$. We insert 
$\Delta\chi^2$ into the cumulative distribution function for 
the $\chi^2$ distribution for one degree of freedom to obtain 
the confidence level (CL) for the \phitwo\ value. The curve
$1\!-\!{\rm CL}$ for all \phitwo\ values is plotted in 
Fig.~\ref{fig:phi2constraint}. From this curve we obtain 
a central value and $1\sigma$ error 
$\phi^{}_2= (87\,\pm 17)^\circ$; the 90\% CL interval around 
the central value is $59^\circ\!<\!\phi^{}_2\!<\!115^\circ$.

In summary, we have measured the branching fraction, longitudinal
polarization fraction $f^{}_L$, and \cp\ asymmetry coefficients 
\arhorho\ and \srhorho\ for \brhorho\ decays.
Our results are
$B=\left[\,24.4\,\pm 2.2\,^{+3.8}_{-4.1}\,\right]\times 10^{-6}$,
$f^{}_L =
0.951\,^{+0.033}_{-0.039}\,^{+0.029}_{-0.031}$,
${\cal A} = 0.00\,\pm 0.30\,^{+0.10}_{-0.09}$, and
${\cal S} = 0.09\,\pm 0.42\,\pm 0.08$,
where the first error listed is statistical and
the second error listed is systematic. 
These results are consistent with previously-published
measurements~\cite{babar_alpha}. The result for $f^{}_L$ is 
consistent with the theoretical expectation based on QCD 
factorization~\cite{kagan}, and the results for
${\cal A}$ and ${\cal S}$ are consistent with 
the case of no \cp\ violation. From an isospin
analysis we obtain the constraint
$59^\circ\!<\!\phi^{}_2\!<\!115^\circ$ at~90\%~CL.

We thank the KEKB group for the excellent operation of the
accelerator, the KEK cryogenics group for the efficient
operation of the solenoid, and the KEK computer group and
the National Institute of Informatics for valuable computing
and Super-SINET network support. We acknowledge support from
the Ministry of Education, Culture, Sports, Science, and
Technology of Japan and the Japan Society for the Promotion
of Science; the Australian Research Council and the
Australian Department of Education, Science and Training;
the National Science Foundation of China under contract
No.~10175071; the Department of Science and Technology of
India; the BK21 program of the Ministry of Education of
Korea and the CHEP SRC program of the Korea Science and
Engineering Foundation; the Polish State Committee for
Scientific Research under contract No.~2P03B 01324; the
Ministry of Science and Technology of the Russian
Federation; the Ministry of Higher Education, 
Science and Technology of the Republic of Slovenia;  
the Swiss National Science Foundation; the National Science Council and
the Ministry of Education of Taiwan; and the U.S.\
Department of Energy.

\end{document}

%% file: author-conf2005.tex
\affiliation{Aomori University, Aomori}
\affiliation{Budker Institute of Nuclear Physics, Novosibirsk}
\affiliation{Chiba University, Chiba}
\affiliation{Chonnam National University, Kwangju}
\affiliation{University of Cincinnati, Cincinnati, Ohio 45221}
\affiliation{University of Frankfurt, Frankfurt}
\affiliation{Gyeongsang National University, Chinju}
\affiliation{University of Hawaii, Honolulu, Hawaii 96822}
\affiliation{High Energy Accelerator Research Organization (KEK), Tsukuba}
\affiliation{Hiroshima Institute of Technology, Hiroshima}
\affiliation{Institute of High Energy Physics, Chinese Academy of Sciences, Beijing}
\affiliation{Institute of High Energy Physics, Vienna}
\affiliation{Institute for Theoretical and Experimental Physics, Moscow}
\affiliation{J. Stefan Institute, Ljubljana}
\affiliation{Kanagawa University, Yokohama}
\affiliation{Korea University, Seoul}
\affiliation{Kyoto University, Kyoto}
\affiliation{Kyungpook National University, Taegu}
\affiliation{Swiss Federal Institute of Technology of Lausanne, EPFL, Lausanne}
\affiliation{University of Ljubljana, Ljubljana}
\affiliation{University of Maribor, Maribor}
\affiliation{University of Melbourne, Victoria}
\affiliation{Nagoya University, Nagoya}
\affiliation{Nara Women's University, Nara}
\affiliation{National Central University, Chung-li}
\affiliation{National Kaohsiung Normal University, Kaohsiung}
\affiliation{National United University, Miao Li}
\affiliation{Department of Physics, National Taiwan University, Taipei}
\affiliation{H. Niewodniczanski Institute of Nuclear Physics, Krakow}
\affiliation{Nippon Dental University, Niigata}
\affiliation{Niigata University, Niigata}
\affiliation{Nova Gorica Polytechnic, Nova Gorica}
\affiliation{Osaka City University, Osaka}
\affiliation{Osaka University, Osaka}
\affiliation{Panjab University, Chandigarh}
\affiliation{Peking University, Beijing}
\affiliation{Princeton University, Princeton, New Jersey 08544}
\affiliation{RIKEN BNL Research Center, Upton, New York 11973}
\affiliation{Saga University, Saga}
\affiliation{University of Science and Technology of China, Hefei}
\affiliation{Seoul National University, Seoul}
\affiliation{Shinshu University, Nagano}
\affiliation{Sungkyunkwan University, Suwon}
\affiliation{University of Sydney, Sydney NSW}
\affiliation{Tata Institute of Fundamental Research, Bombay}
\affiliation{Toho University, Funabashi}
\affiliation{Tohoku Gakuin University, Tagajo}
\affiliation{Tohoku University, Sendai}
\affiliation{Department of Physics, University of Tokyo, Tokyo}
\affiliation{Tokyo Institute of Technology, Tokyo}
\affiliation{Tokyo Metropolitan University, Tokyo}
\affiliation{Tokyo University of Agriculture and Technology, Tokyo}
\affiliation{Toyama National College of Maritime Technology, Toyama}
\affiliation{University of Tsukuba, Tsukuba}
\affiliation{Utkal University, Bhubaneswer}
\affiliation{Virginia Polytechnic Institute and State University, Blacksburg, Virginia 24061}
\affiliation{Yonsei University, Seoul}
  \author{K.~Abe}\affiliation{High Energy Accelerator Research Organization (KEK), Tsukuba} 
  \author{K.~Abe}\affiliation{Tohoku Gakuin University, Tagajo} 
  \author{I.~Adachi}\affiliation{High Energy Accelerator Research Organization (KEK), Tsukuba} 
  \author{H.~Aihara}\affiliation{Department of Physics, University of Tokyo, Tokyo} 
  \author{K.~Aoki}\affiliation{Nagoya University, Nagoya} 
  \author{K.~Arinstein}\affiliation{Budker Institute of Nuclear Physics, Novosibirsk} 
  \author{Y.~Asano}\affiliation{University of Tsukuba, Tsukuba} 
  \author{T.~Aso}\affiliation{Toyama National College of Maritime Technology, Toyama} 
  \author{V.~Aulchenko}\affiliation{Budker Institute of Nuclear Physics, Novosibirsk} 
  \author{T.~Aushev}\affiliation{Institute for Theoretical and Experimental Physics, Moscow} 
% ...
\collaboration{The Belle Collaboration}

%% file: rhorho_preprint.bbl
\begin{thebibliography}{99}

\bibitem{alpha} The naming convention 
$\alpha$ ($=\phi_2$), $\beta$ ($=\phi_1$), and 
$\gamma$ ($=\phi_3$) is also used in the literature.

\bibitem{babar_rho0rho0}
B.\,Aubert {\it et al.\/} (BaBar Collaboration),
Phys.\ Rev.\ Lett.\ {\bf 94}, 131801 (2005).

\bibitem{kekb}
S.~Kurokawa and E.~Kikutani, 
Nucl.\,Instr.\,and\,Meth.\,A {\bf 499}, 1 (2003),
and other papers included in this volume.

\bibitem{belle_detector}
A.~Abashian {\it et al.\/} (Belle Collaboration),
Nucl.\ Instr.\ and Meth.\ A {\bf 479}, 117 (2002).

\bibitem{Ushiroda} Y. Ushiroda (Belle SVD2 Group),
Nucl.\ Instr.\ and Meth.\ A {\bf 511}, 6 (2003). 

\bibitem{tagging}
H.\ Kakuno {\it et al.}, 
Nucl.\ Instr.\ and Meth.\ A {\bf 533}, 516 (2004). 

\bibitem{KSFW}
S.\,H.\,Lee {\it et al.\/} (Belle Collaboration), 
Phys.\ Rev.\ Lett.\ {\bf 91}, 261801 (2003).

\bibitem{argus} 
H.\,Albrecht {\it et al.\/} (ARGUS Collaboration), 
Phys.\ Lett.~B {\bf 241}, 278 (1990).

\bibitem{Long} O.\ Long {\it et al.}, Phys.\ Rev.~D {\bf 68}, 034010 (2003).

\bibitem{pdg} 
S.\ Eidelman {\it et al.\/} (PDG), Phys.\ Lett.~B {\bf 592}, 1 (2004).

\bibitem{gronau_london} M.\ Gronau and D.\ London,
Phys.\ Rev.\ Lett.\ {\bf 65}, 3381 (1990).

\bibitem{falk} A.\ Falk {\it et al.}, 
Phys.\ Rev.~D {\bf 69}, 011502(R) (2004).

\bibitem{babar_alpha} 
B.\,Aubert {\it et al.\/} (BaBar Collaboration),
Phys.\ Rev.\ Lett.\ {\bf 93}, 231801 (2004).

\bibitem{kagan}  A.\ Kagan, Phys.\ Lett.~B {\bf 601}, 151 (2004).

\end{thebibliography}
